# Antiferromagnetic correlations in superconducting YBa$_2$Cu$_3$O$_{6+x}$ samples as seen from optical absorption data; Comparison with the results of neutron and muon experiments

## Vladimir N. Samovarov[*] and Volodymyr L. Vakula

*B. Verkin Institute for Low Temperature Physics and Engineering of the National Academy of Sciences of Ukraine, 47 Lenin Ave., Kharkov 61103, Ukraine*

**Abstract**

Correlation in temperature behavior of the optical, neutron, and muon experimental data is established for YBa$_2$Cu$_3$O$_{6+x}$ in both the pseudogap state at $T_c < T < T^*$ and the superconducting state at $T < T_c$. We have showed that our absorption spectra measured in the vicinity of the optical gap are very sensitive to the development of antiferromagnetic fluctuations and to the formation of a stripe structure. The inelastic neutron scattering and zero-field muon spin relaxation data analyzed in this paper were taken from the measurements made by Dai et al. and Sonier et al., respectively. Comparative analysis of the independent results could be very helpful in understanding the interrelation of superconductivity and antiferromagnetism in cuprate superconductors.

Keywords: optical absorption, YBa$_2$Cu$_3$O$_{6+x}$, antiferromagnetic fluctuations, stripe ordering

## 1. Introduction

Antiferromagnetic (AF) correlations, their relation to the stripe ordering, as well as the coexistence of them and the superconductivity in copper oxides are a point of intensive discussion in the physics of high-$T_c$ superconductivity. The most promising approach to investigation of these problems is to compare the data of various independent experimental techniques. Here we report the results of the comparative analysis performed for YBa$_2$Cu$_3$O$_{6+x}$ on the basis of the data obtained in inelastic neutron scattering [1] and zero-field muon spin relaxation [2] experiments and in our optical interband absorption measurements on variously doped thin YBa$_2$Cu$_3$O$_{6+x}$ films for both normal and superconducting (SC) states. To our knowledge, it is the first time that such an analysis is made for cuprate high-$T_c$ superconductors.

The following facts served as the motivation for this paper. *First*, it is hardly possible to clarify the mechanism of high-$T_c$ superconductivity and to determine the contribution of AF correlations within just one particular tech-nique. It should be noted that the interpretation of the interesting data on AF subsystem of YBa$_2$Cu$_3$O$_{6+x}$ obtained in muon spin relaxation experiments remains ambiguous (see discussion in Refs. [2, 3] and refs. therein). *Second*, as far as we know, other authors took mostly temperature data on the amplitude of the commensurate magnetic resonance (imaginary part of magnetic susceptibility) to make a comparison with the results of other techniques. In Ref. [4], for example, such a comparison was made with the temperature behavior of low-frequency intraband conductivity of YBa$_2$Cu$_3$O$_{6+x}$. Detailed and subtle neutron scattering measurements of the temperature evolution of the momentum- and frequency-integrated resonance intensity (fluctuating magnetic moment associated with the resonance) were made only in Ref. [1]. The resonance data were used there to describe the behavior of the electronic specific heat. In our opinion, the integrated resonance data could be more informative if used more widely for comparative analyses of independent results. It should also be noted that the nature of the resonance remains controversial and is treated within the framework of several different models (see Ref. [5] and refs. therein). *Third*, although the sensitivity of the optical interband transitions to the pseudogap (PG) and SC states has already been shown [6-8], the interrelation between the interband transitions and the AF fluctuations in metallic $YBa_2Cu_3O_{6+x}$ still lacked some direct evidence, and first of all, direct correlation with the neutron data efficiently used for probing the AF subsystem.

---

[*] Corresponding author. Tel.: +38-057-308576; fax: +38-057-322370; e-mail: samovarov@ilt.kharkov.ua

Thus, the existence of the correlated temperature behavior of the results obtained within the three techniques in both normal and SC states was not initially obvious for the cuprate superconductors including $YBa_2Cu_3O_{6+x}$. In this connection, the comparative analysis of the independent results would certainly be of interest.

## 2. Experiment and discussion

We measured the A, A+3J, A+4J, and A+8J absorption bands peaked near and above the optical gap $E_g$=1.7-1.8 eV and their changes upon temperature and doping variation of $YBa_2Cu_3O_{6+x}$ films ($l$=2300 Å).

In the insulator phase ($x$=0.35), the room temperature absorption spectrum $\alpha l(\omega)$ [7, 9, 10] displays three features (see Fig. 1): (i) the boundary A band at 1.77 eV, which may reflect the existence of the Zhang-Rice excitons [11]; (ii) the exciton-bimagnon A+3J band at 2.12 eV, arising from the additional creation of a bimagnon with energy $3J$ (two coupled magnons at the boundary of the Brillouin zone), $J$≈0.12 eV is the AF exchange constant; (iii) the A+8J maximum at ≈2.7 eV, associated with the creation of magnons in the process of relaxation of the excited fermions within their bands. The A and A+3J bands were also observed in the temperature range region of a short-range AF order at $T$=300-400 K > $T_N$. As the temperature was lowered into and within the long-range AF phase (below $T_N$), all the observed maxima became stronger [7, 9, 10]. As it can be seen from Fig. 1, the magnon-assisted peaks are well correlated with the features in the excitation spectrum $R_{2m}$ of the two-magnon $3J$ band observed in the Raman experiments (see Ref. [12] and refs. therein).

As we take films with higher doping level at 300 K, the A and A+3J bands become weaker and are no longer observed in the metallic phase at $x$>0.5-0.6 [7, 9, 10]. Nevertheless, these bands are clearly seen in the PG state at $T<T^*$, and, what is very important, the new A+4J absorption peak appears in the spectra. Fig. 2 shows the difference absorption $\Delta\alpha l$=$\alpha l(T)$-$\alpha l(190K)$ in the frequency region of the A+3J and A+4J bands for the film with $T_c$=74K. It is seen that the A+3J and A+4J peaks become pronounced[1] at 2.14 eV and 2.27 eV, respectively, as the film is cooled below the temperature $T^*$≈160 K which corresponds to the pseudogap transition temperature deduced from other methods including inelastic neutron scattering [1]. The Gaussian dispersion of the peaks is found to be ≈0.05 eV which is very close to the low-temperature value for the A+3J band observed in the AF films. The dispersion is weakly temperature dependent and this implies an increase in the integrated intensities $I_{3J}$ и $I_{4J}$ of the two peaks. A similar picture was observed for the ortho-I and ortho-II films with $T_c$=88K ($T^*$=120K) and 51K, respectively. The appearance of the A and A+3J bands is indicative of the development of an AF order (sublattice fluctuating magnetization). The development of the A+4J peak can be naturally explained in terms of the stripe superstructure formed of alternating insulator and metallic elongated domains with antiphase interdomain boundaries, which is suggested to appear at the same temperature $T^*$ at which the A+4J peak becomes observable. The stripe structure is characterized by a weak interdomain exchange interaction (≈0.1J) and much stronger (≈J) exchange interaction along a stripe [13]. In the stripe scenario, the A+4J peak arises from the excitation of two non-interacting magnons in adjacent AF stripes, while the A+3J peak is due to two coupled magnons within one rather long AF stripe. The optical absorption measurements scale times of $10^{-14}$s at which the dynamical stripe structure can be considered quasistatic.

---
[1] For some films weak traces of these bands were observed in the temperature region above $T^*$.

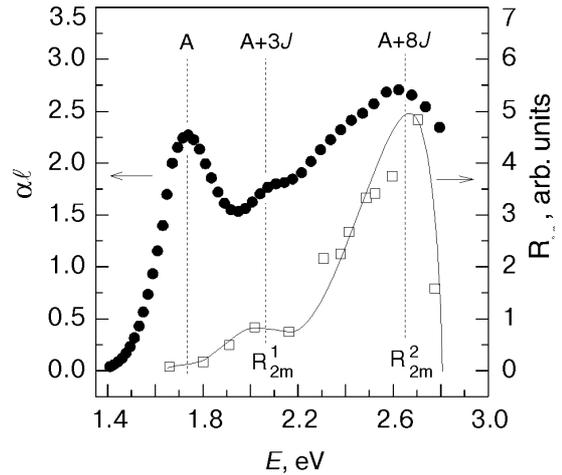

Fig. 1. Solid circles represent the RT optical absorption spectrum $\alpha l(\omega)$ of the insulator $YBa_2Cu_3O_{6+x}$ film with $x$=0.3. Open squares represent the intensity $R_{2m}$ of the Raman scattering two-magnon peak as a function of incident photon energy measured for $YBa_2Cu_3O_6$ at 300 K (see Ref. [12] and references therein). The solid curve is a fit to the Raman data.

Figs. 3 and 4 show the temperature behavior: (i) of the integrated absorption $I_{3J,4J}/I_{3J,4J}(T_c)$; (ii) of the integrated intensity $<m_{res}^2>/<m_{res}^2>(T_c)$ of the magnetic resonance peak at 34 meV in $YBa_2Cu_3O_{6.6}$ ($T_c$≈68 K), where $<m_{res}^2>$ is the mean-squared (fluctuating) moment associated with the resonance; (iii) of the muon depolarization rate $\lambda(T)/\lambda(T_c)$ for $YBa_2Cu_3O_{6.7}$. The data of Fig.3 are presented in the temperature region $T\leq T^*$ that includes both the PG and SC states, Fig. 4 shows the same results on a larger scale in the interval $T_c\leq T\leq T^*$.

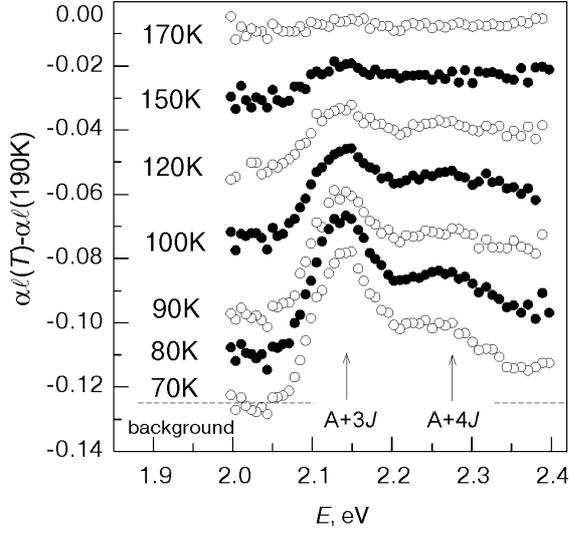

Fig. 2. Development of the A+3J and A+4J absorption peaks on cooling of the metallic YBa₂Cu₃O₆₊ₓ film with $T_c \approx 74K$.

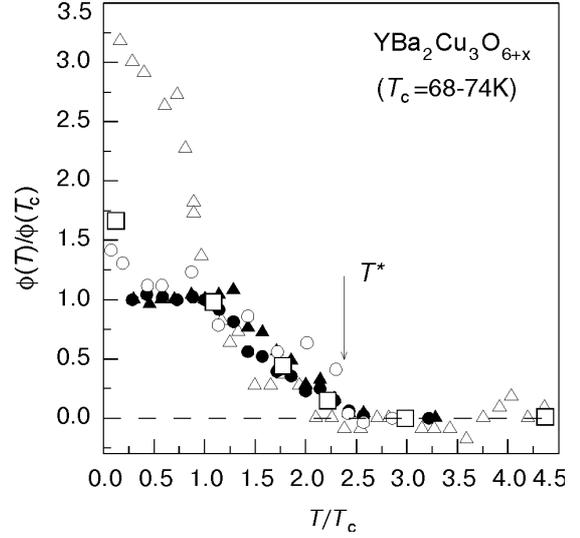

Fig. 3. Normalized temperature changes $\varphi(T)/\varphi(T_c)$, where $\varphi(T)$ is: (i) the integral intensity $I_{3J}(T)$, $I_{4J}(T)$ of the A+3J (solid circles) and A+4J (solid triangles) optical resonances measured in this paper; (ii) the integrated intensity $< m_{res}^2 > (T)$ (open squares) and the peak intensity $S_{res}(q, \omega)$ (open triangles) of the magnetic resonance taken from Ref. [1]; (iii) the muon depolarization rate $\lambda(T)$ according to Ref. [2].

The mean-squared moment $< m_{res}^2 >$ was obtained in Ref. [1] by integrating the resonance peak intensity $S_{res}(q,\omega)$ over the resonance part of the acoustic spectrum.

The muon depolarization rate $\lambda(T)$ was determined in Ref. [2] from the polarization function $G_z(t) = G_z^{KT}(t) \times \exp(-\lambda t)$, where $G_z^{KT}(t)$ is the Kubo-Toyabe function that describes the depolarization of the muon spin by the randomly orientated static magnetic fields of the nuclear dipoles, while the exponential function is due to an additional source of magnetic field. It is supposed that the exponential function can reflect the existence of fluctuating magnetic moments associated with the atomic spins and $\lambda(T) \propto \nu^{-1}(T)$, where $\nu(T)$ is the frequency of spin fluctuations at muon site. From this viewpoint it could be quite naturally to attribute the increase in $\lambda(T)$ observed on cooling of the samples to a growth of the fluctuating AF moment. A similar increase in $\lambda(T)$ was reported in the papers on AF order in borides [14], on small AF clusters in the heavy-fermion system URu₂Si₂ [15], on dynamics of the stripe structure in La₂₋ₓSrₓCu₁₋ᵧZnᵧO₄ [16]. But, when speaking of YBa₂Cu₃O₆₊ₓ [3], it is hardly possible to interpret the behavior of $\lambda(T)$ in terms of AF dynamical Cu spin correlations since the Cu spin fluctuation rate is too fast ($10^{10}$ to $10^{12}$ s⁻¹) to be detected with muon spin relaxation experiments.

Figs. 3 and 4 clearly demonstrate that the three independent sets of data are in rather good agreement in both PG and SC states. At $T \leq T^*$, the results satisfy the scaled dependence

$$\frac{I_{3J,4J}}{I_{3J,4J}(T_c)} \approx \frac{< m_{res}^2 >}{< m_{res}^2 > (T_c)} \approx \frac{\lambda(T)}{\lambda(T_c)} = f\left(\frac{T}{T_c}\right). \qquad (1)$$

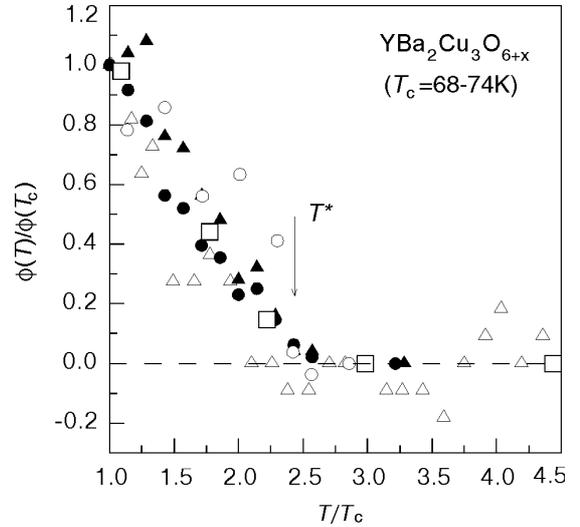

Fig. 4. The same data as in Fig. 3 on a larger scale for $T > T_c$.

The observed difference between the peak and integrated magnetic resonance intensities below $T_c$ (Fig. 3) clearly illustrates the resonance narrowing in the SC state reported in Ref. [1].

Analysis of the optical and neutron data suggests that the growth of $f(T/T_c)$ reflects the enhancement of AF coherence and the development of a stripe structure at $T^* \leq (2.25 \div 2.5)T_c$. We can suppose that the muon data can be interpreted in the same way. The stripe structure persists in the SC phase at $T<T_c$, $f(T/T_c)\approx$const (within the experimental errors).

The established correlation in behavior of the optical, neutron, and muon data in the PG and SC states clearly demonstrates that the optical absorption features near $E_g$ are able to provide information about the development of AF fluctuations in metallic samples.

## 3. Conclusions

1. As the temperature is lowered, the exciton A band, the exciton-bimagnon A+3J band, and the exciton-two-magnon A+4J band appear in the absorption spectra of the metallic (SC) $YBa_2Cu_3O_{6+x}$ films with both weak and nearly optimal doping. The bands become clearly observable in the PG temperature region at $T<T^*$. Their appearance can be attributed to the development of the stripe structure. According to this scenario, the A and A+3J absorption peaks, which are typical of an antiferromagnetically ordered $YBa_2Cu_3O_{6+x}$ insulator, are closely related to the AF stripes. The A+4J component, which can be clearly seen only below $T^*$, is also magnetic in origin. Its nature can be naturally explained by the excitation of two non-interacting magnons in the adjacent AF stripes separated by a thin metallic domain wall. We do not exclude that a certain contribution to the spectral weight of these bands can be also from multimagnon interactions and/or from strong quantum fluctuations.

2. It is shown that in the temperature region $T<T^*$ including the SC state, there is a rather good correlation between the temperature behavior of the A+3J and A+4J bands, the temperature evolution of the mean-squared fluctuating moment associated with the magnetic resonance peak and the dynamical muon spin relaxation rate. The analysis of the scaled function (1) shows a growth of the AF correlations as well as the development of the stripe structure on cooling of the samples. The temperature of the phase segregation is found to be $T^*\approx2.5T_c$ for the samples with $T_c\approx70$K and $T^*\approx1.5T_c$ for the samples with $T_c\approx85$K. The AF correlations and the stripe ordering remain in the SC state.

3. The A, A+3J, and A+4J optical absorption bands near and above the charge transfer gap can be actively used for probing the stripe ordering in both normal and SC states of cuprate superconductors. The optical interband spectroscopy is proved to be quite sensitive to the AF correlations. In this connection, it would be of particular interest to see the effect of magnetic field on the bands (magnetic sublattice collinearity) and to carry out absorption (reflectivity) experiments with polarized light which could provide some new important information about the nature of the stripe and SC states.

The more detailed discussion of the obtained results will be published in Low Temp. Phys. 29 (November/December 2003).